\documentclass{PoS}
\usepackage{xcolor}
\usepackage{color,colortbl}
\usepackage{wrapfig}

\newcommand{\FullStop}{.}

\newcommand{\fm}{\textrm{fm}}

\newcommand{\Bu}{\ensuremath{B^+}}

\newcommand{\Vub}{\ensuremath{V_{ub}}}

\newcommand{\spose}[1]{\hbox to 0pt{#1\hss}}

\newcommand{\inapprox}{\mathrel{\spose{\lower 3pt\hbox{$\mathchar"218$}}
 \raise 2.0pt\hbox{$\mathchar"232$}}}

\newcommand{\chpt}{\raise0.4ex\hbox{$\chi$}PT}
\newcommand{\schpt}{S\chpt}

\title{The $D_s$, $D^+$, $B_s$ and $B$ decay constants from $2+1$ flavor lattice QCD}

\ShortTitle{The $D_s$, $D^+$, $B_s$ and $B$ decay constants from $2+1$ flavor lattice QCD}

\newcommand{\SU}[1]{\ensuremath{^{#1}}}
\newcommand{\BNL}{a}
\newcommand{\WUSTL}{b}
\newcommand{\OSU}{c}
\newcommand{\UTAH}{d}
\newcommand{\UIUC}{e}
\newcommand{\SAIC}{f}
\newcommand{\GRAN}{g}
\newcommand{\IU}{h}
\newcommand{\APS}{i}
\newcommand{\AZ}{j}
\newcommand{\FNAL}{k}
\newcommand{\UGLA}{l}
\newcommand{\SYR}{m}
\newcommand{\CUB}{n}
\newcommand{\RIKENBNL}{o}
\newcommand{\UCSB}{p}

\author{%
A.~Bazavov\SU\BNL,
C.~Bernard\SU\WUSTL,
C.~Bouchard\SU\OSU,
C.~DeTar\SU\UTAH,
D.~Du\SU\UIUC,
A.X.~El-Khadra\SU\UIUC,
J.~Foley\SU\UTAH,
E.D.~Freeland\SU{\SAIC},
E.~G\'{a}miz\SU\GRAN,
Steven~Gottlieb\SU{\IU},
U.M.~Heller\SU\APS,
J.~Kim\SU\AZ,
J.~Komijani\SU\WUSTL,
A.S.~Kronfeld\SU\FNAL,
J.~Laiho\SU{\UGLA,\SYR},
L.~Levkova\SU\UTAH,
P.B.~Mackenzie\SU\FNAL,
D.~Mohler\SU\FNAL,
E.T.~Neil\SU{\FNAL,\CUB,\RIKENBNL,\alpha},
M.B.~Oktay\SU\UTAH,
S.~Qiu\SU\UTAH,
\speaker{J.N.~Simone}\SU{,\FNAL,\beta},
R.L.~Sugar\SU\UCSB,
D.~Toussaint\SU\AZ,
R.S.~Van~de~Water\SU\FNAL,
and
R.~Zhou\SU{\IU,\FNAL} \\
\llap{\SU\BNL}Physics Department, Brookhaven National Laboratory, Upton NY 11973, USA \\
\llap{\SU\WUSTL}Department of Physics, Washington University, St.~Louis, MO 63130, USA \\
\llap{\SU\OSU}Department of Physics, The Ohio State University, Columbus, OH 43210, USA \\
\llap{\SU\UTAH}Physics Department, University of Utah, Salt Lake City, UT 84112, USA \\
\llap{\SU\UIUC}Physics Department, University of Illinois, Urbana, IL 61801, USA \\
\llap{\SU\SAIC}Liberal Arts Department, School of the Art Institute of Chicago, Chicago, IL 60603, USA \\
\llap{\SU\GRAN}CAFPE and Depto. de F\'{i}sica Te\'{o}rica y del Cosmos, Universidad de Granada, Granada, Espa\~{n}a \\
\llap{\SU\IU}Department of Physics, Indiana University, Bloomington, IN 47405, USA \\
\llap{\SU\APS}American Physical Society, One Research Road, Ridge, NY 11961, USA \\
\llap{\SU\AZ}Department of Physics, University of Arizona, Tucson, AZ 85721, USA \\
\llap{\SU\FNAL}Fermi National Accelerator Laboratory, Batavia, IL 60510, USA \\
\llap{\SU\UGLA}SUPA, School of Physics and Astronomy, University of Glasgow, Glasgow, G12~8QQ, UK \\
\llap{\SU\SYR}Department of Physics, Syracuse University, Syracuse, NY 13244, USA \\
\llap{\SU\CUB}Department of Physics, University of Colorado, Boulder, CO 80309, USA \\
\llap{\SU\RIKENBNL}RIKEN-BNL Research Center, Brookhaven National Laboratory, Upton, NY 11973, USA \\
\llap{\SU\UCSB}Department of Physics, University of California, Santa Barbara, CA 93106, USA \\
Email:\hspace{1.0em} \llap{\SU\alpha}\email{ethan.neil@colorado.edu} \hspace{2.0em}   \llap{\SU\beta}\email{simone@fnal.gov}
}
\author{Fermilab Lattice and MILC Collaborations}

\abstract{We present a study of the $D$ and $B$ leptonic decay
constants on the MILC $N_f=2+1$ asqtad gauge ensembles using
asqtad-improved staggered light quarks and clover heavy quarks
in the Fermilab interpretation.  Our previous analysis \cite{Bazavov:2011aa}
computed the decay constants at lattice spacings
$a \approx 0.14, 0.11$ and $0.083 \,\fm$.
We have extended the
simulations to finer
$a \approx 0.058$ and $0.043 \,\fm$
lattice spacings, and have also increased statistics;
this allows us to address many important
sources of uncertainty.
Technical advances include a two-step two-point fit procedure, better tuning of the heavy quark
masses and a better determination of the axial-vector current matching.
The present analysis remains blinded, so here
we focus on the improvements and their predicted impact
on the error budget compared to the prior analysis.
}

\FullConference{31st International Symposium on Lattice Field Theory LATTICE 2013\\
		 July 29 -- August 3, 2013\\
		 Mainz, Germany}

\definecolor{colMC}{rgb}{1.00,0.39,0.28} 
\definecolor{colCs}{rgb}{1.00,0.65,0.00} 
\definecolor{colFn}{rgb}{0.60,0.80,0.20} 
\definecolor{colSF}{rgb}{0.68,0.85,0.90} 
\definecolor{colUF}{rgb}{0.93,0.51,0.93} 

\newcommand{\TABensembles}{
\begin{wraptable}{r}{0.55\textwidth}
\begin{center}{\tiny
\begin{tabular}{lllllllrr} \hline\hline
\textbf{id} &\textbf{$\mathbf{a}$ [fm]} &$\textbf{beta}$ &$\mathbf{m_l/m_h}$ &$\mathbf{am_h}$ &$\mathbf{m_h/m_s}$ &$\mathbf{r_1/a}$ &$\mathbf{N_\mathit{config}}$ &$\mathbf{N_\mathit{tsrc}}$ \\ \hline
\rowcolor{colUF}
A    &0.043  &7.81     &0.2   &0.014  &1.079 &7.208 & 801  & 4 \\
\rowcolor{colSF}
B    &0.059  &7.46     &0.1   &0.018  &1.019 &5.307  & 827  & 4 \\
\rowcolor{colSF}
C    &0.058  &7.465    &0.139 &0.018  &1.024 &5.330 & 801  & 4 \\
\rowcolor{colSF}
D    &0.058  &7.47     &0.2   &0.018  &1.028 &5.353 & 673  & 8 \\
\rowcolor{colSF}
E    &0.058  &7.48     &0.4   &0.018  &1.037 &5.399 & 593  & 4 \\
\rowcolor{colFn}
F    &0.083  &7.075    &0.05  &0.031  &1.255 &3.738 & 791  & 4 \\
\rowcolor{colFn}
G    &0.083  &7.08     &0.1   &0.031  &1.256 &3.755 &1015  & 4 \\
\rowcolor{colFn}
H    &0.083  &7.085    &0.15  &0.031  &1.262 &3.772 & 984  & 4 \\
\rowcolor{colFn}
I    &0.082  &7.09     &0.2   &0.031  &1.267 &3.789 &1931  & 4 \\
\rowcolor{colFn}
J    &0.081  &7.11     &0.4   &0.031  &1.290 &3.858 &1996  & 4 \\
\rowcolor{colCs}
K    &0.11   &6.76     &0.1   &0.05   &1.489 &2.739 &2099  & 4 \\
\rowcolor{colCs}
L    &0.11   &6.76     &0.14  &0.05   &1.489 &2.739 &2110  & 4 \\
\rowcolor{colCs}
M    &0.11   &6.76     &0.2   &0.05   &1.489 &2.739 &2259  & 4 \\
\rowcolor{colCs}
N    &0.11   &6.79     &0.4   &0.05   &1.534 &2.821 &2052  & 4 \\
\rowcolor{colMC}
O    &0.14   &6.572    &0.2  &0.0484 &1.156 &2.222  & 631  &24 \\
\hline
\end{tabular}}
\end{center}
\caption{MILC asqtad ensembles and parameters.}
\label{tab:MILCensembles}
\end{wraptable}
}

\newcommand{\FIGchiralFitD}{
\begin{figure}
\begin{center}
\includegraphics[height=0.97\textheight]{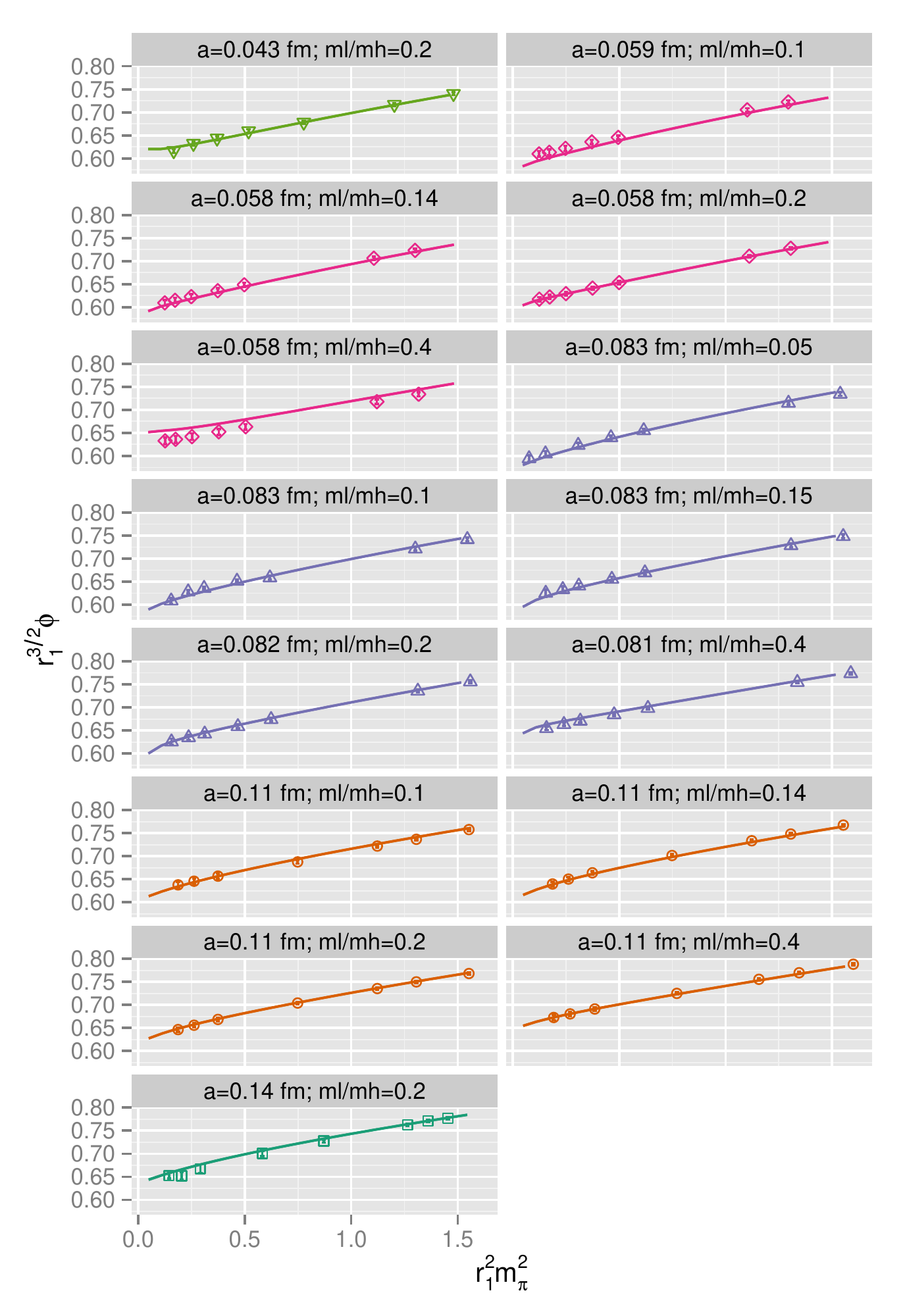}
\end{center}
\caption{Fit of all $D$ system partially quenched data to the $\schpt$ model function.}
\label{fig:DSXPTfit}
\end{figure}
}

\newcommand{\FIGchiralFitB}{
\begin{figure}
\begin{center}
\includegraphics[height=0.97\textheight]{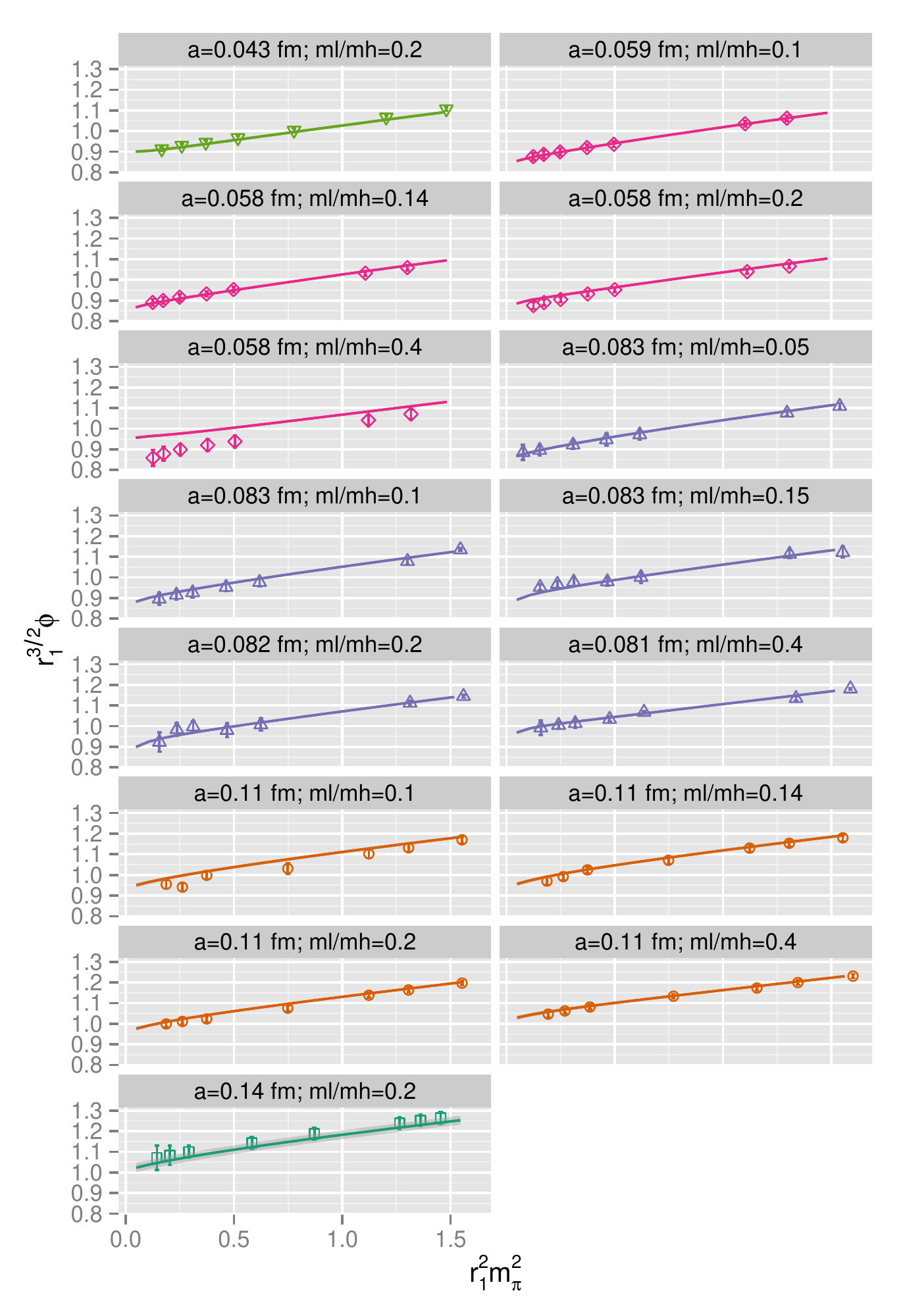}
\end{center}
\caption{Fit of all $B$ system partially quenched data to the $\schpt$ model function.}
\label{fig:BSXPTfit}
\end{figure}
}

\newcommand{\FIGchiralExtrapDB}{
\begin{figure}\renewcommand{\arraystretch}{0.8}
\begin{center}
\begin{tabular}{c}
\includegraphics[width=0.71\textwidth]{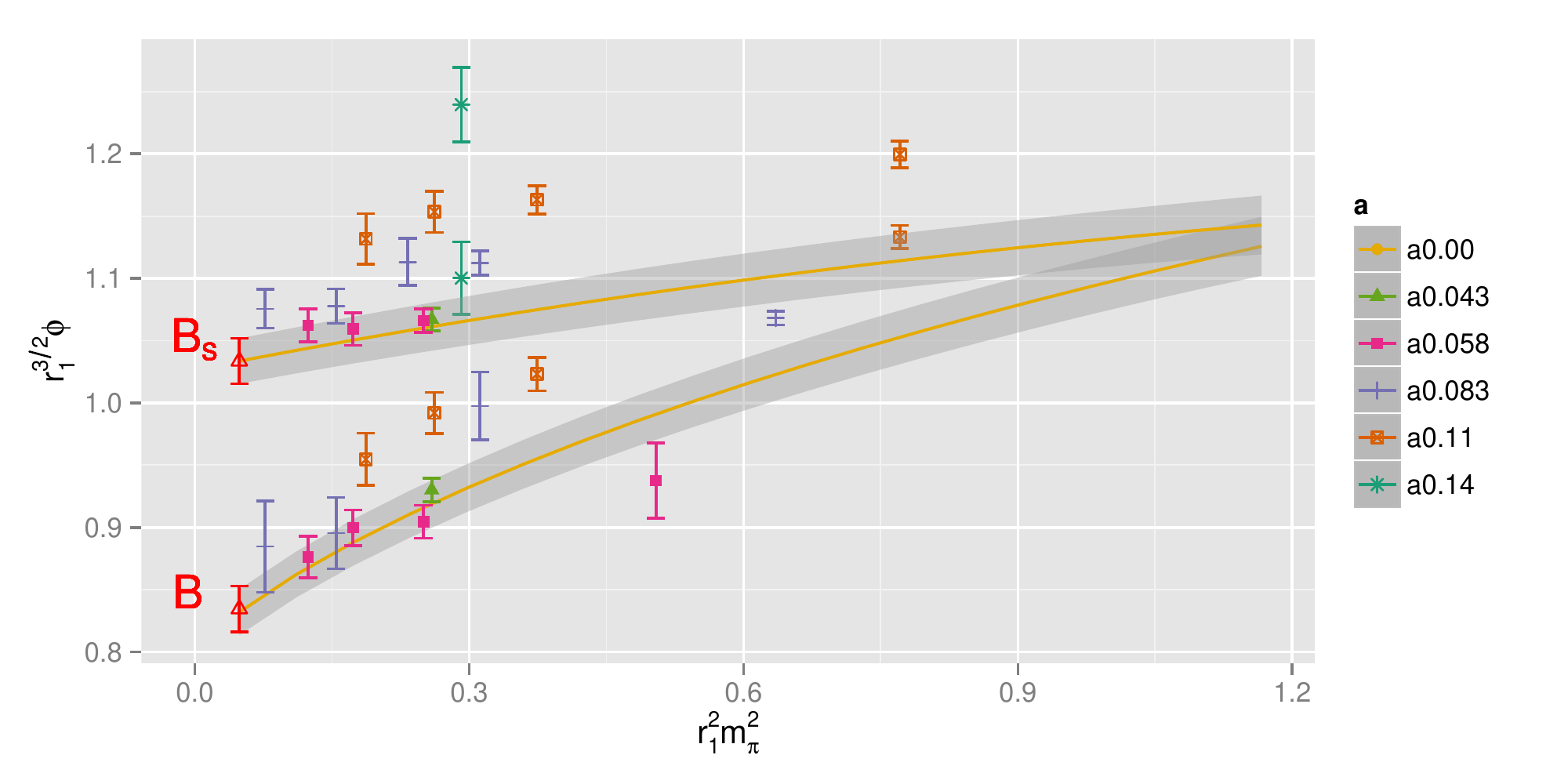} \\
\includegraphics[width=0.71\textwidth]{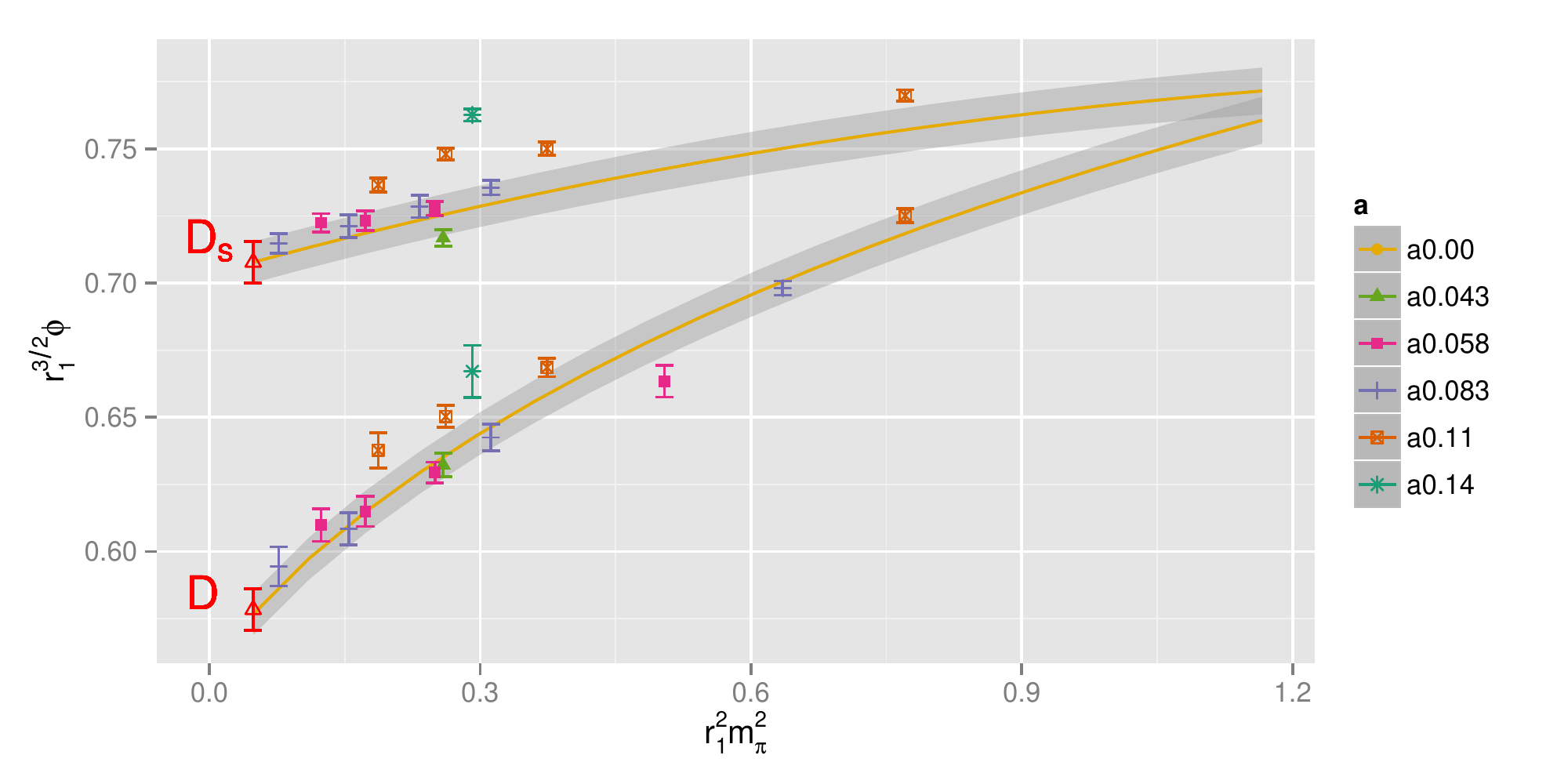}
\end{tabular}
\end{center}
\caption{Combined chiral continuum extrapolation for the $B$ (top) and $D$ (bottom) systems.
The curves and statistical error bands in the $a\to0$ limit at physical quark masses are labeled ``a0.00''.
The extrapolated values (and errors) are the points labeled by triangles at the physical pion mass.
Note: these results are
blinded by normalization factors known only to a few collaboration members, who are outside the analysis group.
Note: the subset of points shown for each of the $B_s$ and $D_s$ extrapolations serve merely as a guide
since their valence masses only approximate the physical strange quark mass: $r_1|m_q-m_s|<0.07$.
}
\label{fig:bothSXPTextrap}
\end{figure}
}

\newcommand{\FIGerrorBudgetProjected}{
\begin{figure}
\begin{center}
\includegraphics[width=0.80\textwidth]{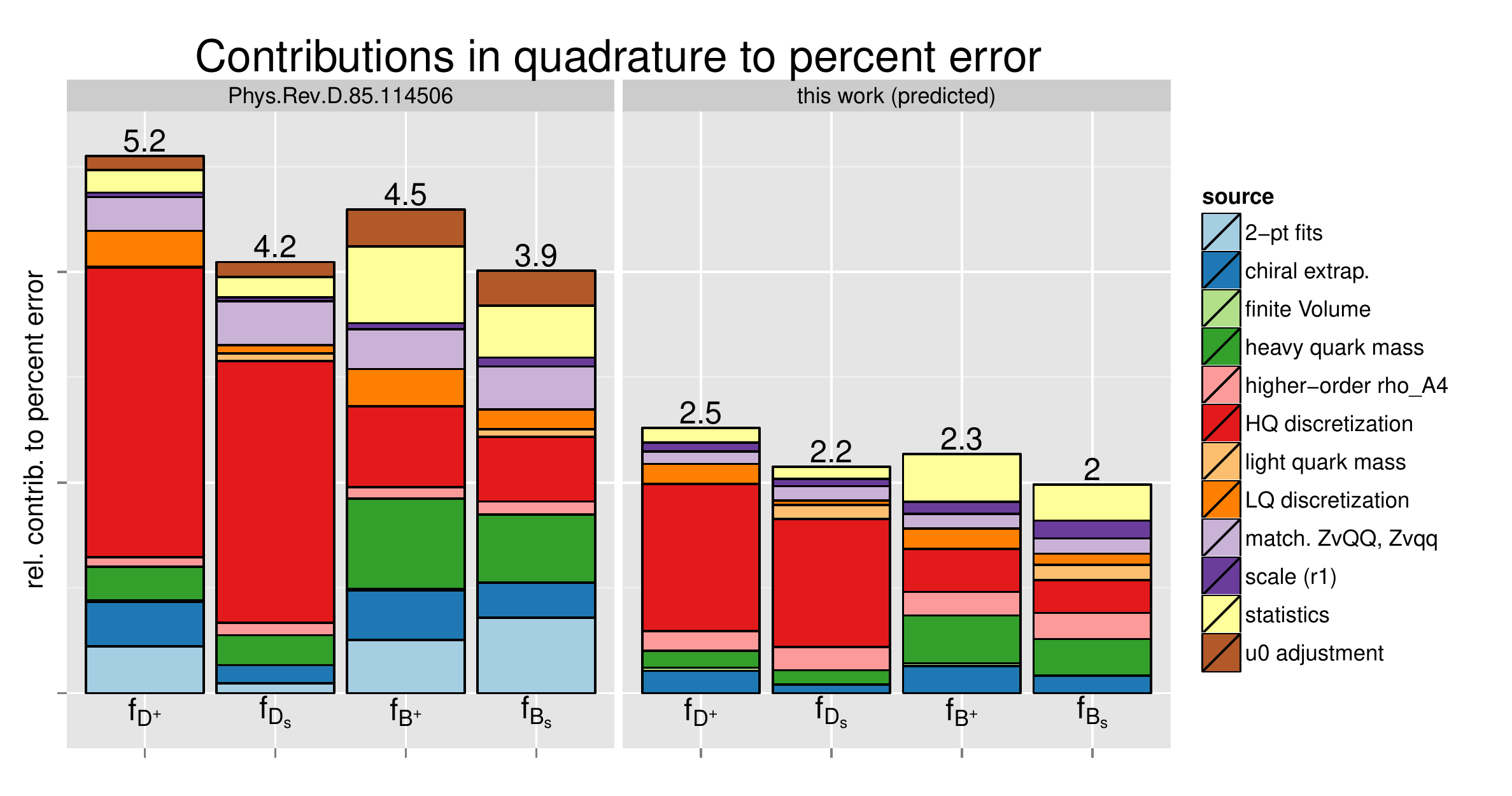}
\end{center}
\vspace*{-2.0ex}
\caption{The relative contributions to the error in quadrature from sources of uncertainty.
On the left a summary of our previous analysis \cite{Bazavov:2011aa}. On the right
are our predictions for this analysis.}
\label{fig:ErrorBudget}
\end{figure}

}

\begin{document}

\section{Introduction}

Decays of $B$ and $D$ mesons provide an important testing ground for the CKM paradigm for flavor-changing weak interactions in the Standard Model.  The $D$ and $B$ decay constants, which encapsulate the role of QCD interactions in these decay processes, are a crucial theoretical input.  Precise calculation of these strong-coupling quantities, enabled by lattice simulations, is very important in predicting experimental rates for rare decays, such as the process $B^0_{(s)}\to\mu^+\mu^-$ observed recently by LHCb and CMS \cite{CMSandLHCbCollaborations:2013pla}, where the $B$-meson decay constants enter into the Standard Model rates. In addition, precise knowledge of the $\Bu$-decay constant in combination with the observed $\Bu\to\tau^+\nu$ decay rate probes the $V$-$A$ structure of the
$W u b$ vertex and helps in understanding the tension between inclusive and exclusive determinations of $|\Vub|$.


\TABensembles
This study uses simulations on the ensembles listed in
Table~\ref{tab:MILCensembles}.
Since Ref.~\cite{Bazavov:2011aa} we have added ensembles labeled A
through E, at two finer lattice spacings, and ensemble F at a sea-quark mass
$m_l=0.05 m_h$, nearer the physical mass. Statistical accuracy is also better,
with about $3.6\times$ more $N_\mathit{config}\cdot N_\mathit{tsrc}$ combinations than in Ref.~\cite{Bazavov:2011aa}.
Better statistics, finer lattice spacings and a nearly physical sea quark mass all
help to control the leading
systematic uncertainties observed in Ref.~\cite{Bazavov:2011aa} that arise
from heavy- and light-quark discretization effects
and the (chiral) extrapolation.
This analysis also benefits from reduced uncertainty from the input charm
and bottom quark masses, due to a retuning of the masses with improved
techniques and higher statistics, and a reanalysis of the nonperturbative
matching of the flavor conserving heavy and light vector currents, again
with higher statistics.

\section{Two-point fits}

We use a two-stage procedure for performing the two-point fits.
In the first stage, plots of the effective mass are inspected for a stable ``plateau" at large values of the source-sink separation $t$.  A range $[t_{p,{min}}, t_{p,{max}}]$ is then chosen based on the correlator signal-to-noise ratio (SNR), maintaining SNR $\geq 10$ for all correlators and holding the SNR range approximately fixed over different ensembles; this translates to fits at approximately equal physical distances on the different lattice spacings.  The first-stage fit is carried out to a standard two-state functional form (one oscillating), with no excited states included.

For the second stage, the results of the first-stage fit are used to set empirical Bayesian priors; best-fit values give the prior means, and the width is set equal to the best-fit one-sigma error estimate times an inflation factor of $3$, to ensure that the second-stage fit parameters are not overconstrained.  We have tested that increasing the inflation factor beyond this point has negligible effect on the resulting final classical error estimates; however, the use of this two-stage procedure serves to stabilize the fits performed over bootstrap resampled data, by reducing the occurrence of outliers.  We fit a basis of four (or five) smeared and local source two-point correlators including the
two correlators having an $\mathcal{O}(a)$-improved axial-current at the sink in stage two.
We find good isolation of the ground state when including four to five
states (plus an equal number of oscillating states) and fitting
down to $t_{{min}}=2$.

\section{Chiral fits}

The decay constant $f_{H_q}$ for a meson $H_q$ is related to $\phi_{H_q}= f_{H_q} \sqrt{M_{H_q}}$, where
\begin{equation}
r_1^{3/2}\phi_{H_q} =
\left(\frac{r_1}{a}\right)^{3/2} \sqrt{Z_{V^4_{QQ}}\,Z_{V^4_{qq}}}\; \rho_{A^4_{Qq}}\;\; a^{3/2} \phi_{H_q}^{\mathit{lat}}
\quad\FullStop
\end{equation}
Values for $r_1/a$ are shown in Table~\ref{tab:MILCensembles}.
The flavor-conserving vector-current renormalization factors
$Z_{V^4_{QQ}}$ and $Z_{V^4_{qq}}$ are found nonperturbatively, while
$\rho_{A^4_{Qq}}=1+\mathcal{O}(\alpha_V)$ is known to one-loop order and is near unity.
The $a^{3/2}\phi_{H_q}^{\mathit{lat}}$ are determined by fitting two-point functions.

Guided by heavy meson staggered chiral perturbation theory \cite{Aubin:2005aq}, we fit to
the function
\begin{equation}
\phi_{H_q}\left(m_q,m_h,m_l\right)=\Phi_H\left[1 + \Delta f_{H_q}\left(m_q,m_h,m_l\right) +
                                    P\left(m_q,m_h,m_l\right) +
                                    K\left(am_Q\right) + c_a a^2 \right]
\qquad\FullStop
\label{Eqn:ModelFun}
\end{equation}
Term $\Delta f_{H_q}$, parameterizing the NLO chiral
logarithms, includes hyperfine splitting effects. They are
corrected for staggered taste effects at finite lattice spacing and for finite volume.
The polynomial $P$ includes analytic terms up
to second order in the quark masses. The $K$ terms, which parametrize the
leading-order heavy quark discretization effects, are constrained in fits according to power counting
estimates \cite{Bazavov:2011aa}. The residual heavy-quark discretization
error for the physical $\phi_{H_q}$ is then incorporated into the overall statistical error.

\FIGchiralFitD
\FIGchiralFitB
Figure~\ref{fig:DSXPTfit} shows a preliminary fit for
the model in Equation~(\ref{Eqn:ModelFun}) with all the $D$-meson
$\phi_{H_q}$ simulation results at the five lattice spacings listed in
Table~\ref{tab:MILCensembles}.
Figure~\ref{fig:BSXPTfit} is the corresponding plot for the $B$-meson system.
The figures indicate that the model adequately represents
the simulation results. Points differing only by valence light-quark mass
are shown in the individual plot subpanels.
A reasonable fit is obtained even when the fit curve may visibly deviate
from simulation results on a particular ensemble since points there are strongly correlated.

\FIGchiralExtrapDB
Figure~\ref{fig:bothSXPTextrap} shows the combined
continuum and chiral extrapolation curve
(and error band) for the $D$- and $B$-meson systems.
Shown overlaying each of the extrapolated $B$ and $D$ fit curves are the (finite $a$) ``full QCD'' points
where $m_q=m_l$.
No points correspond exactly to the tuned strange quark mass, consequently,
we show a subset of points with roughly $m_q\approx m_s$ overlaying each of the $B_s$ and $D_s$ curves.

\section{Predicted errors and outlook}

Our results remain blinded, hence, we do not quote
values for the decay constants here. We will continue to use
the blinded analysis to understand systematic
effects. In Figure~\ref{fig:ErrorBudget} we compare
predicted errors in this study to Ref.~\cite{Bazavov:2011aa}.
This analysis has higher statistics and includes results at
finer $a\approx0.058$ and $0.043\,\fm$ lattice spacings
leading to improved estimates for discretization
effects modeled in the chiral fit function. We anticipate this will
lead to a reduction in the residual discretization errors for the physical decay
constants.  The addition of results with $m_l=0.05 m_h$, nearer to the
physical quark (ensemble F in Table~\ref{tab:MILCensembles}), narrows the
extent of the chiral extrapolation, which
is expected to reduce the residual chiral extrapolation
uncertainty.

\FIGerrorBudgetProjected

The current study includes several technical improvements compared to
Ref.~\cite{Bazavov:2011aa}.  We have introduced a two-step procedure for
two-point fitting which allows us to stably model more excited states
and, hence, better utilize two-point data at small times where the
signal-to-noise ratio is larger. This fit procedure
also better preserves expected (significant) correlations
among points by reducing the likelihood of finding
outliers in output bootstrap distributions that tend to wash out correlations.
The new procedure and better statistics help to
directly reduce statistical errors as well as the ``2-pt fit''
uncertainty previously estimated from plausible variations in
fitting procedures.
We have reduced the error in the decay constants due to the input heavy quark mass through
improvements to the tuning process used to determine the charm
and bottom quark masses: a) We have four times the statistics than in
prior tuning runs. b) We employ priors in energy-momentum
dispersion relation fits which help stabilize the kinetic masses that
are matched to the physical $D_s$ and $B_s$ masses. c) We compensate
for mistuning of the strange sea-quark
masses in simulations. d) We smoothly extend the charm and bottom tunings from the
subset of ensembles used for tuning to all other ensembles.
Uncertainties in the decay constants due to the flavor-conserving matching factors
$Z_{V^4}$ for both clover and staggered currents
have been reduced by new determinations using
better stochastic color wall meson sources
and higher statistics.

From Figure~\ref{fig:ErrorBudget}, we anticipate that
the total error in quadrature for the decay constants in this study 
will be around half of the error found in Ref.~\cite{Bazavov:2011aa}.
We are continuing to refine this analysis before unblinding
the values for the decay constants.

\noindent\textbf{Acknowledgments} \\
This work was supported by the U.S. Department of Energy, the National Science Foundation, and the URA 
Visiting Scholars' Program. Fermilab and BNL are operated under contracts De-AC02-07CH11359 and DE-AC02-98CH10886 respectively,
with the DOE.
Computations were carried out at the Argonne Leadership Computing Facility,
the National Center for Atmospheric Research,
the National Center for Supercomputing Resources,
the National Energy Resources Supercomputing Center,
the National Institute for Computational Sciences, 
the Texas Advanced Computing Center, and
the USQCD facilities at Fermilab, under grants from the NSF and DOE.

\bibliographystyle{JHEP}
\bibliography{bib/fDfB}

\end{document}